\documentclass{desyproc}

\newcommand{\lwig}{\mbox{\;\raisebox{.3ex}
    {$<$}$\!\!\!\!\!$\raisebox{-.9ex}{$\sim$}\;}}
\newcommand{\gwig}{\mbox{\;\raisebox{.3ex}
    {$>$}$\!\!\!\!\!$\raisebox{-.9ex}{$\sim$}}\;}

\usepackage{bm}
\usepackage{epsfig}
\usepackage{citesort}
\usepackage{graphicx}

\begin{document}
\title{Cosmological axion bounds}

\author{{\slshape Steen Hannestad$^1$, Jan Hamann$^1$, Alessandro Mirizzi$^2$, Georg G.~Raffelt$^3$, Yvonne Y.~Y.~Wong$^4$}\\[1ex]
$^1$~Department of Physics and Astronomy,
 University of Aarhus, DK-8000 Aarhus C, Denmark \\
 $^2$~Institut f\"ur theoretische Physik, Universit\"at Hamburg,
Luruper Chaussee 149, D-22761 Hamburg, Germany \\
 $^3$~Max-Planck-Institut f\"ur Physik (Werner-Heisenberg-Institut), F\"ohringer Ring 6, D-80805 M\"unchen, Germany \\
$^4$~Institut f\"ur Theoretische Physik E, RWTH Aachen University, D-52056 Aachen, Germany}
\contribID{lindner\_axel}

\desyproc{DESY-PROC-2009-05}
\acronym{Patras 2009} 
\doi  

\maketitle

\begin{abstract}
We discuss current cosmological constraints on axions, as well as future sensitivities. Bounds on axion hot dark matter are discussed first, and subsequently we discuss both current and
future sensitivity to models in which axions play the role as cold dark matter, but where the Peccei-Quinn symmetry is not restored during reheating.
\end{abstract}

The Peccei-Quinn (PQ) mechanism provides a simple explanation for the smallness of the QCD $\Theta$ parameter~\cite{Peccei:2006as}.
A consequence of this is the existence of axions, low mass pseudoscalars similar to pions, except that their mass and coupling strength are suppressed by a factor $f_\pi/f_a$, where $f_\pi \simeq 93$ MeV and $f_a$ is the PQ scale.
The axion mass is given by the relation
\begin{equation}
m_a = \frac{z^{1/2}}{1+z} \frac{f_\pi m_\pi}{f_a} = \frac{6 \, {\rm eV}}{f_a/10^6 \, {\rm GeV}},
\end{equation}
where $z \equiv m_u/m_d \sim 0.3-0.6$.
Thus, there is a tight relation between $m_a$ and $f_a$, known as the axion line.
Axions couple to photons with a coupling $g_{a \gamma}$ of order
$g_{a \gamma} \sim \frac{\alpha}{f_a}$
and are therefore in principle detectable even if $f_a$ is much larger than the electroweak energy scale. A larger number of direct detection experiments use this coupling to photons to search for axions, and are described in detail elsewhere in these proceedings.

Astrophysics also provides a stringent bound on the axion-photon coupling (see \cite{Raffelt:2006cw} for a thorough discussion). The most restrictive bound comes from constraints on the horizontal branch (HB) lifetime of globular cluster stars. If an additional source of energy loss from the core is present, the core Helium burning phase can be shortened to a point where the predicted number of HB stars in a globular cluster is in conflict with observations. The bound from this argument roughly corresponds to $g_{a \gamma} \lwig 10^{-10} \, {\rm GeV}^{-1}$.

While this bound is formally very stringent, it is also model dependent, and it is possible to construct models with an axion photon coupling much smaller than the normally predicted $\alpha/f_a$. However, in this case cosmology provides an important lower bound on $f_a$ coming from the unavoidable coupling of axions to quarks. In this case, the main thermalisation mechanism is axion-pion conversion, $a \pi \leftrightarrow \pi \pi$. Provided that $f_a \lwig {\rm few} \times 10^7 \, {\rm GeV}$ axions couple sufficiently strongly to thermalise completely after the QCD phase transition at $T \sim 150$ MeV. If this is the case axions automatically provide a source of hot dark matter because they will have masses in the eV range.

\begin{figure}[h!]
\hspace{25mm}
\includegraphics[width=8.cm]{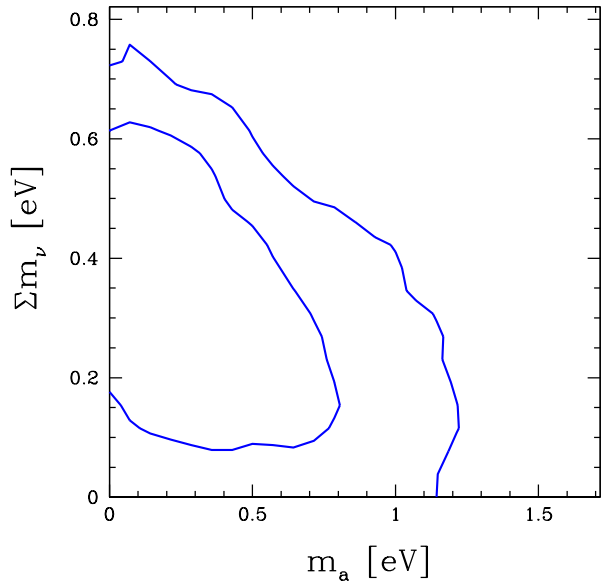}
\caption{68\% and 95\% contours in the $\sum m_\nu$-$m_a$
  plane (taken from Ref.~\cite{Hannestad:2008js}). 
\label{fig:contours}}
\end{figure}

Therefore, cosmological constraints on light neutrinos can be also be applied to axions in this range. The current upper bound on the axion mass is of order 0.5-1 eV \cite{Hannestad:2005df,Hannestad:2008js}, corresponding to $f_a \gwig 10^7 \, {\rm GeV}$. Fig.~\ref{fig:contours}, taken from \cite{Hannestad:2008js}, shows the bound on $m_a$ and $\sum m_\nu$ simultaneously.

If the PQ scale is much higher, axions never thermalise in the early universe, and their presence is caused solely by non-thermal production.
During the QCD epoch of the early universe, a non-thermal
mechanism produces axions as nonrelativistic coherent field
oscillations that can play the role of cold dark
matter~\cite{Sikivie:2006ni}. In terms of the initial ``misalignment
angle'' $\Theta_{\rm i}=a_{\rm i}/f_{a}$ relative to the
CP-conserving minimum of the axion potential, the cosmic axion
density is~\cite{Bae:2008ue}
\begin{equation}
 \omega_a = \Omega_{a}h^2\simeq 0.195\,\Theta_{\rm i}^2
 \left(\frac{f_{a}}{10^{12}~{\rm GeV}}\right)^{1.184}\,.
 \label{eq:omega}
\end{equation}
If $\Theta_{\rm i}^2$ is of order unity, axions
provide the dark matter of the universe if $f_{a}\sim 10^{12}$~GeV
($m_{a}\sim10~\mu{\rm eV}$).

One may also consider axions in another range beyond the classical
cosmological window. In a scenario where the PQ symmetry is not
restored during or after inflation, a single value
$-\pi<\Theta_{\rm i}<+\pi$ determines the axion density in our Hubble volume. It is
possible that $\Theta_{\rm i}\ll1$, allowing for $f_{a}\gg
10^{12}$~GeV. This ``anthropic axion window'' is motivated because the PQ mechanism
presumably is embedded in a greater framework. In particular, the PQ
symmetry emerges naturally in many string scenarios, where $f_a$ is naturally high
(see also \cite{Pi:1984pv}).

An interesting signature of such a high $f_a$ is the presence of
primordial isocurvature
fluctuations that can show up in future data.
When axions acquire a mass during the QCD
epoch, axion field fluctuations from the de Sitter expansion during inflation become dynamically relevant in the form of
isocurvature fluctuations that are uncorrelated with the adiabatic
fluctuations inherited by all other matter and radiation from the
inflaton field. The isocurvature amplitude depends on both $f_{a}$ and $H_{\rm I}$, the Hubble parameter during inflation, so
observational limits on isocurvature fluctuations exclude certain
regions in this parameter space
\cite{Lyth:1989pb,Komatsu:2008hk}.
Since there is no trace of isocurvature fluctuations in existing
data, perhaps a more interesting question is the remaining window
for axions to show up in future data.

Now going into slightly more detail, when
the PQ symmetry breaks at some large temperature $T\sim v_{\rm PQ}$,
the relevant Higgs field will settle in a minimum corresponding to
$\Theta_{\rm i}=a_{\rm i}/f_{a}$, where $-\pi\leq\Theta_{\rm
  i}\leq+\pi$.  We assume that this happens before cosmic inflation,
so throughout our observable universe we have the same initial
condition except for fluctuations imprinted by inflation itself.
The cosmic energy density in axions is given by Eq.~(\ref{eq:omega}) with $\Theta_{\rm i}^2$ replaced by
$\langle \Theta^2\rangle_{\rm i}=\Theta_{\rm i}^2+\sigma_\Theta^2$, where
$\sigma_\Theta^2 = H_I^2/(4 \pi^2 f_{a}^2)$ is the inflation-induced variance, with $H_I$ the Hubble parameter
during inflation.
All cosmologically viable models have $\Theta_{\rm i}^2 \gg \sigma_\Theta^2$.
Assuming that all of the cold dark matter consists of
axions, according to current cosmological data $\omega_{\rm
  a}=\omega_{\rm c} = 0.109\pm0.004$. Assuming $\sigma_\Theta$ is small, one finds
$\Theta_{\rm i}=0.748\, \left(\frac{10^{12}~{\rm GeV}}{f_{a}}\right)^{0.592}$
as a unique relationship between the initial misalignment angle and
the axion decay constant.

As mentioned, axion-induced isocurvature fluctuations are uncorrelated with the
adiabatic fluctuations inherited by other matter and radiation
components from the inflaton, and the isocurvature fraction, $\alpha$, of the total
fluctuation power spectrum is given by \cite{Hamann:2009yf}
\begin{equation}
\alpha \simeq  7.5\times10^{-3}
\left(\frac{2.4 \times 10^{-9}}{A_{\rm S}} \right)
\left(\frac{0.109}{\omega_{\rm c}} \right) \left(
\frac{H_I}{10^{7} \ {\rm GeV}}\right)^2  \left( \frac{10^{12} \ {\rm
GeV}}{f_{a}} \right)^{0.816} \label{eq:axionalpha},
\end{equation}
where $A_{\rm S}= {\cal P}(k=k_0)$ is the amplitude of the total
primordial scalar power spectrum at the pivot scale $k_0=0.002~{\rm Mpc}^{-1}$.

\cite{Hamann:2009yf} considered several different current and future data sets in order to constrain $\alpha$:
1) Current data: WMAP plus auxiliary data sets. 2) Planck: Simulated $TT$, $TE$ and $EE$ spectra up to
    $\ell = 2000$ from the Planck satellite \cite{planck}. 3) CVL: Simulated, noiseless $TT$, $TE$ and $EE$
    spectra up to $\ell = 2000$. Roughly equivalent to the projected CMBPol experiment \cite{Baumann:2008aq}.

Data set 1 gives $\alpha < 0.09$ at 95\% confidence,
consistent with the findings of Komatsu et
al.~\cite{Komatsu:2008hk}.
For the future experiments and if no isocurvature signal shows up,
we forecast 95\%-credible upper limits of
$\alpha < 0.042$ for Planck and $\alpha<0.017$ for CVL.

\begin{figure}[h!]
\hspace{25mm}
\includegraphics[width=9cm]{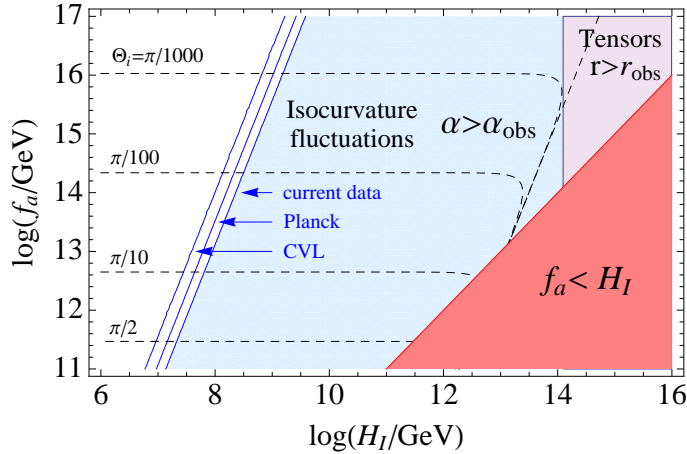}
\caption{Exclusion and sensitivity regions in the plane of $H_I$
(Hubble rate during inflation) and $f_{a}$ (axion decay
constant), assuming axions are all of the dark matter. The
isocurvature exclusion region based on current data is shown in
light blue. The sensitivity forecasts for
Planck and CVL  are also
indicated. The dashed lines indicate the required $\Theta_{\rm i}$
for a given $f_{a}$
to obtain the full amount of axion dark matter. We also show the
region of excessive tensor modes and the region $f_{a}<H_I$
where our late-inflation scenario is not applicable. \label{fig:tegmark}}
\end{figure}

The constraints and sensitivity
forecasts on the isocurvature fraction $\alpha$ can be translated into axion
parameters using equation~(\ref{eq:axionalpha}). It can we
written in the form
\begin{equation}
H_I=3.5\times10^7~{\rm GeV}
\left(\frac{\alpha}{0.09}\right)^{1/2}
\left(\frac{\omega_{\rm c}}{0.109}\right)^{1/2}
\left(\frac{f_{a}}{10^{12}~{\rm GeV}}\right)^{0.408},
\end{equation}
where the present upper bound on $\alpha$ has been used as a benchmark.
Assuming axions are the dark matter, this constraint is shown in
Fig.~\ref{fig:tegmark} with a line marked $\alpha=0.09$.
In this plot, taken from \cite{Hamann:2009yf}, we also show the relationship between
$f_{a}$ and $\Theta_{\rm i}$ as dashed lines. Future
sensitivities to $\alpha$ from Planck and CVL are
shown labelled with the appropriate $\alpha$ values.

In conclusion, axions have a number of potentially very important consequences for cosmology, depending entirely on the value of the PQ symmetry breaking scale.

For low values of the breaking scale, axions are thermalised in the early universe and can act as a hot dark matter component. Such scenarios can be constrained
by current observations of large scale structure, and while these constraints are formally much less stringent than astrophysical bounds using the axion-photon
coupling, they are also less model dependent.

At intermediate scales, $f_a \sim 10^{12}$ GeV, axions may naturally act as a CDM candidate.

At the other end of the scale, a very high value of the PQ scale may also provide axion cold dark matter, but in this case there may be additional observational
signatures in the form of axion isocurvature fluctuations. Such an isocurvature component could plausible be observed by future CMB experiments, and would provide
a very interesting new window on both axion physics and early universe cosmology.

\begin{footnotesize}

\end{footnotesize}


\begin{thebibliography}{99}


\bibitem{Peccei:2006as}
  R.~D.~Peccei,
  Lect.\ Notes Phys.\  {\bf 741} (2008) 3;
  J.~E.~Kim and G.~Carosi,
  Rev.\ Mod.\ Phys., to be published (arXiv:0807.3125)


\bibitem{Sikivie:2006ni}
  P.~Sikivie,
  Lect.\ Notes Phys.\  {\bf 741} (2008) 19.

\bibitem{Bae:2008ue}
  K.~J.~Bae, J.~H.~Huh and J.~E.~Kim,
  JCAP {\bf 0809} (2008) 005.


\bibitem{Raffelt:2006cw}
  G.~G.~Raffelt,
  Lect.\ Notes Phys.\  {\bf 741}, 51 (2008).

%
\bibitem{Hannestad:2005df}
  S.~Hannestad, A.~Mirizzi and G.~Raffelt,
  JCAP {\bf 0507} (2005) 002;
  A.~Melchiorri, O.~Mena and A.~Slosar,
  Phys.\ Rev.\  D {\bf 76} (2007) 041303.

\bibitem{Hannestad:2008js}
  S.~Hannestad, A.~Mirizzi, G.~G.~Raffelt and Y.~Y.~Y.~Wong,
  JCAP {\bf 0804} (2008) 019.


\bibitem{Pi:1984pv}
  So-Young~Pi,
  Phys.\ Rev.\ Lett.\  {\bf 52} (1984) 1725;
  A.~D.~Linde,
  Phys.\ Lett.\  B {\bf 201} (1988) 437;
  M.~Tegmark, A.~Aguirre, M.~Rees and F.~Wilczek,
  Phys.\ Rev.\  D {\bf 73} (2006) 023505.

\bibitem{Lyth:1989pb}
  D.~H.~Lyth,
  Phys.\ Lett.\  B {\bf 236} (1990) 408;
  M.~Beltr\'an, J.~Garc{\'\i}a-Bellido and J.~Lesgourgues,
  Phys.\ Rev.\  D {\bf 75} (2007) 103507;
  M.~P.~Hertzberg, M.~Tegmark and F.~Wilczek,
  Phys.\ Rev.\  D {\bf 78} (2008) 083507;
  L.~Visinelli and P.~Gondolo,
  Phys.\ Rev.\  D {\bf 80}, 035024 (2009).

\bibitem{Komatsu:2008hk}
  E.~Komatsu {\it et al.}  [WMAP Collaboration],
  Astrophys.\ J.\ Suppl.\  {\bf 180} (2009) 330.


\bibitem{Baumann:2008aq}
  D.~Baumann {\it et al.}  [CMBPol Study Team Collaboration],
  AIP Conf.\ Proc.\  {\bf 1141}, 10 (2009).

\bibitem{planck}
  M.~Bersanelli {\it et al.}, eds., [Planck Science Team],
  arXiv:astro-ph/0604069.

\bibitem{Hamann:2009yf}
  J.~Hamann, S.~Hannestad, G.~G.~Raffelt and Y.~Y.~Y.~Wong,
  JCAP {\bf 0906}, 022 (2009).

\end{thebibliography}
\end{document}